\documentclass[twocolumn,aps,prl,showpacs,superscriptaddress,groupedaddress]{revtex4}
\usepackage[latin9]{inputenc}
\usepackage{fancyhdr}
\usepackage{amsmath}
\usepackage{amssymb}
\usepackage{graphicx}
\usepackage{esint}

\makeatletter
\@ifundefined{textcolor}{}
{%
 \definecolor{BLACK}{gray}{0}
 \definecolor{WHITE}{gray}{1}
 \definecolor{RED}{rgb}{1,0,0}
 \definecolor{GREEN}{rgb}{0,1,0}
 \definecolor{BLUE}{rgb}{0,0,1}
 \definecolor{CYAN}{cmyk}{1,0,0,0}
 \definecolor{MAGENTA}{cmyk}{0,1,0,0}
 \definecolor{YELLOW}{cmyk}{0,0,1,0}
}

\usepackage{endnotes}

\makeatother

\begin{document}

\title{Linear Readout of Object Manifolds }

\author{SueYeon Chung}

\affiliation{Program in Applied Physics, School of Engineering and Applied Sciences,
\\
 Harvard University, Cambridge, MA 02138, USA}

\affiliation{Center for Brain Science, Harvard University, Cambridge, MA 02138,
USA}

\author{Daniel D. Lee}

\affiliation{Department of Electrical and Systems Engineering, University of Pennsylvania,
Philadelphia, PA 19104, USA}

\author{Haim Sompolinsky}

\thanks{Correspondence (haim@fiz.huji.ac.il)}

\affiliation{Center for Brain Science, Harvard University, Cambridge, MA 02138,
USA}

\affiliation{Racah Institute of Physics, Hebrew University, Jerusalem 91904, Israel}

\affiliation{Edmond and Lily Safra Center for Brain Sciences, Hebrew University,
Jerusalem 91904, Israel\smallskip{}
 }

\begin{abstract}
Objects are represented in sensory systems by continuous manifolds
due to sensitivity of neuronal responses to changes in physical features
such as location, orientation, and intensity. What makes certain sensory
representations better suited for invariant decoding of objects by
downstream networks? We present a theory that characterizes the ability
of a linear readout network, the perceptron, to classify objects from
variable neural responses. We show how the readout perceptron capacity
depends on the dimensionality, size, and shape of the object manifolds
in its input neural representation. 
\end{abstract}
\maketitle
High-level perception in the brain involves classifying or identifying
objects which are represented by continuous manifolds of neuronal
states in all stages of sensory hierarchies \cite{dicarlo2007untangling,pagan2013signals,alemi2013multifeatural,bizley2013and,meyers2015intelligent,schwarzlose2008distribution,gottfried2010central}
Each state in an object manifold corresponds to the vector of firing
rates of responses to a particular variant of physical attributes
which do not change object's identity, e.g., intensity, location,
scale, and orientation. It has been hypothesized that object identity
can be decoded from high level representations, but not from low level
ones, by simple downstream readout networks \cite{hung2005fast,dicarlo2007untangling,pagan2013signals,freiwald2010functional,cadieu2014deep,kobatake1994neuronal,rust2010selectivity,schwarzlose2008distribution}.
A particularly simple decoder is the perceptron, which performs classification
by thresholding a linear weighted sum of its input activities \cite{minsky1987perceptrons,gardnerEPL}.
However, it is unclear what makes certain representations well suited
for invariant decoding by simple readouts such as perceptrons. Similar
questions apply to the hierarchy of artificial deep neural networks
for object recognition \cite{serre2005object,goodfellow2009measuring,ranzato2007unsupervised,bengio2009learning,cadieu2014deep}.
Thus, a complete theory of perception requires characterizing the
ability of linear readout networks to classify objects from variable
neural responses in their upstream layer.

A theoretical understanding of the perceptron was pioneered by Elizabeth
Gardner who formulated it as a statistical mechanics problem and analyzed
it using replica theory \cite{gardner1988space,engel2001statistical,advani2013statistical,brunel2004optimal,sompolinsky1990learning,opper1991generalization,rubin2010theory,amit1989perceptron,monasson1992properties}.
In this work, we generalize the statistical mechanical analysis and
establish a theory of linear classification of manifolds synthesizing
statistical and geometric properties of high dimensional signals.
We apply the theory to simple classes of manifolds and show how changes
in the dimensionality, size, and shape of the object manifolds affect
their readout by downstream perceptrons.

\paragraph*{Line segments: }

\begin{figure}
\noindent \begin{centering}
\includegraphics[width=8.5cm]{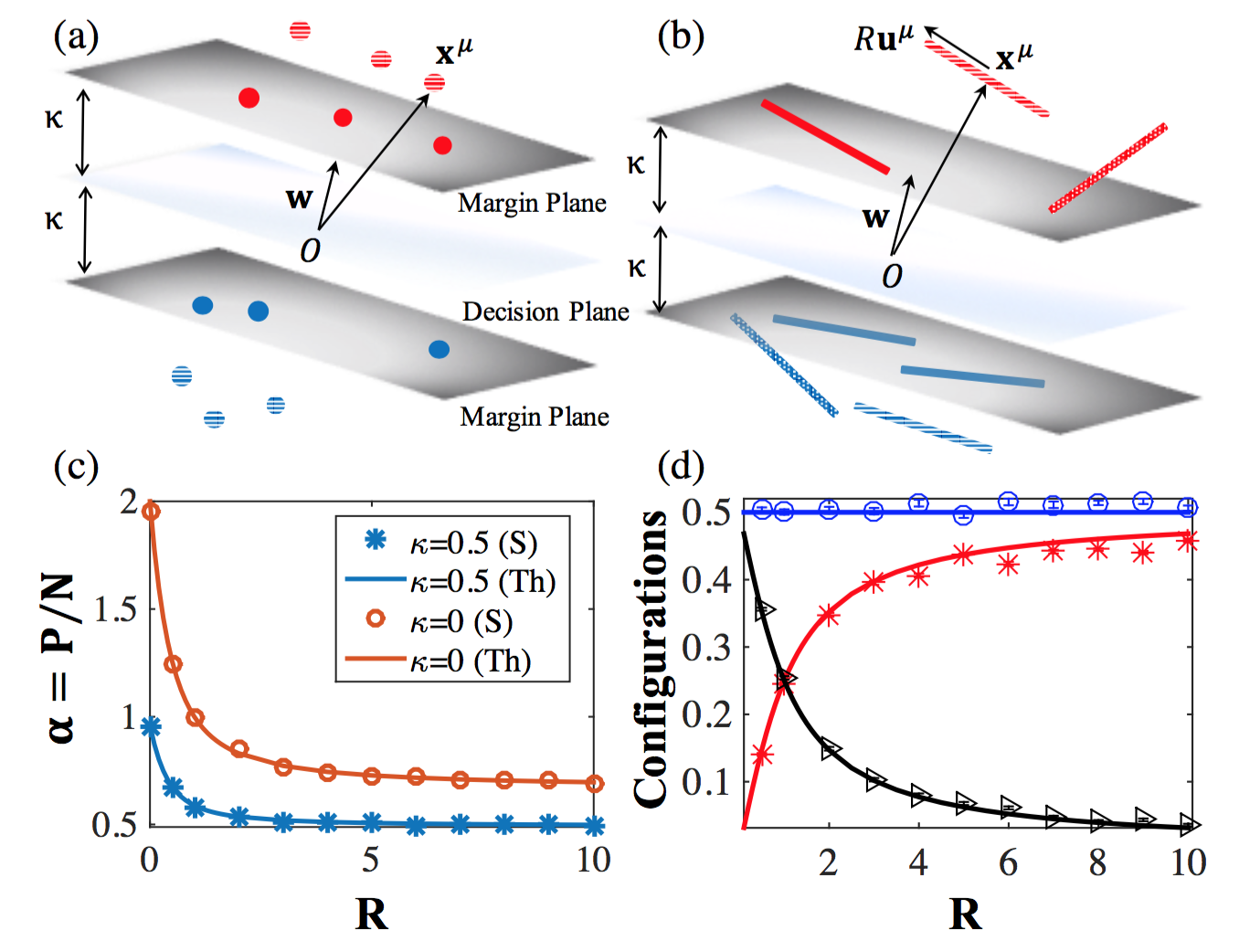} 
\par\end{centering}

\caption{(a) Linear classification of points. (solid) points on the margin,
(striped) internal points. (b) Linear classification of line segments.
(solid) lines embedded in the margin, (dotted) lines touching the
margin, (striped) interior lines. (c) Capacity $\alpha=P/N$ of a
network $N=200$ as a function of $R$ with margins $\kappa=0$ (red)
and $\kappa=0.5$ (blue). Theoretical predictions (lines) and numerical
simulation (markers, see SM for details) are shown. (d) Fraction of
different line configurations at capacity with $\kappa=0$. (red)
lines in the margin, (blue) lines touching the margin, (black) internal
lines. \label{fig:PerceptronLines} }
\end{figure}

One-dimensional object manifolds arise naturally from variation of
stimulus intensity, such as visual contrast, which leads to approximate
linear modulation of the neuronal responses of each object. We model
these manifolds as line segments and consider classifying $P$ such
segments in $N$ dimensions, expressed as $\left\{ \mathbf{x}^{\mu}+Rs\mathbf{u}^{\mu}\right\} $,
$-1\le s\le1$, $\mu=1,...,P$. The $N$-dimensional vectors $\mathbf{x}^{\mu}\in\mathcal{R}^{N}$
and $\mathbf{u}^{\mu}\in\mathcal{R}^{N}$ denote respectively, the
\emph{centers} and \emph{directions }of\emph{ }the $\mu$-th segment,
and the scalar $s$ parameterizes the continuum of points along the
segment. The parameter $R$ measures the extent of the segments relative
to the distance between the centers (Fig. \ref{fig:PerceptronLines}).

We seek to partition the different line segments into two classes
defined by binary labels $y^{\mu}=\pm1$ . To classify the segments,
a weight vector $\mathbf{w}\in\mathcal{R}^{N}$ must obey $y^{\mu}\mathbf{w}\cdot\left(\mathbf{x}^{\mu}+Rs\mathbf{u}^{\mu}\right)\ge\kappa$
for all $\mu$ and $s$. The parameter $\kappa\ge0$ is known as the
margin; in general, a larger $\kappa$ indicates that the perceptron
solution will be more robust to noise and display better generalization
properties \cite{vapnik1998statistical}. Hence, we are interested
in maximum margin solutions, i.e., weight vectors $\mathbf{w}$ that
yield the maximum possible value for $\kappa$. Since line segments
are convex, only the endpoints of each line segment need to be checked,
namely $\min\,h_{0}^{\mu}\pm Rh^{\mu}=h_{0}^{\mu}-R\left|h^{\mu}\right|\ge\kappa$
where $h_{0}^{\mu}=||\mathbf{w}||^{-1}y^{\mu}\mathbf{w}\cdot\mathbf{x}^{\mu}$
are the fields induced by the centers and $h^{\mu}=||\mathbf{w}||^{-1}y^{\mu}\mathbf{w}\cdot\mathbf{u}^{\mu}$
are the fields induced by the line directions.

\paragraph*{Replica theory:}

The existence of a weight vector $\mathbf{w}$ that can successfully
classify the line segments depends upon the statistics of the segments.
We consider random line segments where the components of $\mathbf{x}^{\mu}$
and $\mathbf{u}^{\mu}$ are i.i.d. Gaussians with zero mean and unit
variance, and random binary labels $y^{\mu}$. We study the thermodynamic
limit where the dimensionality $N\rightarrow\infty$ and number of
segments $P\rightarrow\infty$ with finite $\alpha=P/N$ and $R$.
Following Gardner \cite{gardner1988space} we compute the average
of $\log V$ where $V$ is the volume of the space of perceptron solutions:
\begin{equation}
V=\int_{\left\Vert \mathbf{w}\right\Vert ^{2}=N}d^{N}\mathbf{w}\:\prod_{\mu=1}^{P}\Theta\left(h_{0}^{\mu}-R\left|h^{\mu}\right|-\kappa\right).\label{eq:Volume}
\end{equation}
$\Theta(x)$ is the Heaviside step function. According to replica
theory, the fields are described as sums of random Gaussian fields
$h_{0}^{\mu}=t_{0}^{\mu}+z_{0}^{\mu}$ and $h^{\mu}=t^{\mu}+z^{\mu}$
where $t_{0}$ and $t$ are quenched components arising from fluctuations
in the input vectors $\mathbf{x}^{\mu}$ and $\mathbf{u}^{\mu}$ respectively,
and the $z_{0}$, $z$ fields represent the variability in $h_{0}^{\mu}$
and $h^{\mu}$ resulting from different solutions of $\mathbf{w}$.
These fields must obey the constraint $z_{0}+t_{0}-R\left|z+t\right|\ge\kappa.$
The capacity function $\alpha_{1}(\kappa,R)$ (the subscript $1$
denotes the dimensionality of the manifolds) describes for which $P/N$
ratio the perceptron solution volume shrinks to a unique weight vector.
The reciprocal of the capacity is given by the replica symmetric calculation
(details provided in supplementary materials, SM): 
\begin{equation}
\alpha_{1}^{-1}(\kappa,R)=\left\langle \min_{z_{0}+t_{0}-R\left|z+t\right|\ge\kappa}\frac{1}{2}\left[z_{0}^{2}+z^{2}\right]\right\rangle _{t_{0},t}\label{eq:alphaLinesAverage}
\end{equation}
where the average is over the Gaussian statistics of $t_{0}$ and
$t$. To compute Eq. \eqref{eq:alphaLinesAverage}, three regimes
need to be considered. First, when $t_{0}$ is large enough so that
$t_{0}>\kappa+R\left|t\right|$, the minimum occurs at $z_{0}=z=0$
which does not contribute to the capacity. In this regime, $h_{0}^{\mu}>\kappa$
and $h^{\mu}>0$ implying that neither of the two segment endpoints
reach the margin. In the other extreme, when $t_{0}<\kappa-R^{-1}|t|$,
the minimum is given by $z_{0}=\kappa-t_{0}$ and $z=-\left|t\right|$,
i.e. $h_{0}^{\mu}=\kappa$ and $h^{\mu}=0$ indicating that both endpoints
of the line segment lie on the margin planes. In the intermediate
regime where $\kappa-R^{-1}\left|t\right|<t_{0}<\kappa+R\left|t\right|$,
$z_{0}=\kappa-t_{0}+R|z+t$|, i.e., $h_{0}^{\mu}-R|h^{\mu}|=\kappa$
but $h_{0}^{\mu}>\kappa$, corresponding to only one of the line segment
endpoints touching the margin. In this regime, the solution is given
by minimizing the function $(R\left|z+t\right|+\kappa-t_{0})^{2}+z^{2}$
with respect to $z$. Combining these contributions, we can write
the perceptron capacity of line segments: 
\begin{eqnarray}
\alpha_{1}^{-1}(\kappa,R) & = & \int_{-\infty}^{\infty}Dt\int_{\kappa-R^{-1}|t|}^{\kappa+R|t|}Dt_{0}\frac{\left(R\left|t\right|+\kappa-t_{0}\right)^{2}}{R^{2}+1}\nonumber \\
 & + & \int_{-\infty}^{\infty}Dt\int_{-\infty}^{\kappa-R^{-1}|t|}Dt_{0}\left[(\kappa-t_{0})^{2}+t^{2}\right]\quad\quad\label{eq:alphaCLine}
\end{eqnarray}

with integrations over the Gaussian measure, $Dx\equiv\frac{1}{\sqrt{2\pi}}e^{-\frac{1}{2}x^{2}}dx$.
It is instructive to consider special limits. When $R\rightarrow0,$
Eq. \eqref{eq:alphaCLine} reduces to $\alpha_{1}(\kappa,0)=\alpha_{0}(\kappa)$
where $\alpha_{0}(\kappa)$ is Gardner's original capacity result
for perceptrons classifying $P$ points (the subscript $0$ stands
for zero-dimensional manifolds) with margin $\kappa$ \ref{fig:PerceptronLines}-(a).
Interestingly, when $R=1$, then $\alpha_{1}(\kappa,1)=\frac{1}{2}\alpha_{0}(\kappa/\sqrt{2})$.
This is because when $R=1$ there are no statistical correlations
between the line segment endpoints and the problem becomes equivalent
to classifying $2P$ random points with average norm $\sqrt{2N}$
.

Finally, when $R\rightarrow\infty$, the capacity is further reduced:
$\alpha_{1}^{-1}(\kappa,\infty)=\alpha_{0}^{-1}(\kappa)+1$. This
is because when $R$ is large, the segments become unbounded lines.
In this case, the only solution is for $\mathbf{w}$ to be orthogonal
to all $P$ line directions. The problem is then equivalent to classifying
$P$ center points in the $N-P$ null space of the line directions,
so that at capacity $P=\alpha_{0}(\kappa)(N-P)$.

We see this most simply at zero margin, $\kappa=0$. In this case,
Eq. \eqref{eq:alphaCLine} reduces to a simple analytic expression
for the capacity: $\alpha_{1}^{-1}(0,R)=\frac{1}{2}+\frac{2}{\pi}\arctan R$
(SM). The capacity is seen to decrease from $\alpha_{1}(0,R=0)=2$
to $\alpha_{1}(0,R=1)=1$ and $\alpha_{1}(0,R=\infty)=\frac{2}{3}$
for unbounded lines. We have also calculated analytically the distribution
of the center and direction fields $h_{0}^{\mu}$ and $h^{\mu}$ \cite{abbott1989universality}.
The distribution consists of three contributions, corresponding to
the regimes that determine the capacity. One component corresponds
to line segments fully embedded in these planes. The fraction of these
manifolds is simply the volume of phase space of $t$ and $t_{0}$
in the last term of Eq. \eqref{eq:alphaCLine}. Another fraction,
given by the volume of phase space in the first integral of \eqref{eq:alphaCLine}
corresponds to line segments touching the margin planes at only one
endpoint. The remainder of the manifolds are those interior to the
margin planes. Fig. \ref{fig:PerceptronLines} shows that our theoretical
calculations correspond nicely with our numerical simulations for
the perceptron capacity of line segments, even with modest input dimensionality
$N=200$. Note that as $R\rightarrow\infty$, half of the manifolds
lie in the plane while half only touch it; however, the angles between
these segments and the margin planes approach zero in this limit.
As $R\rightarrow0$ , half of the points lie in the plane \cite{abbott1989universality}.

\paragraph*{$D$-dimensional balls:}

\begin{figure}
\noindent \begin{centering}
\includegraphics[width=8.5cm]{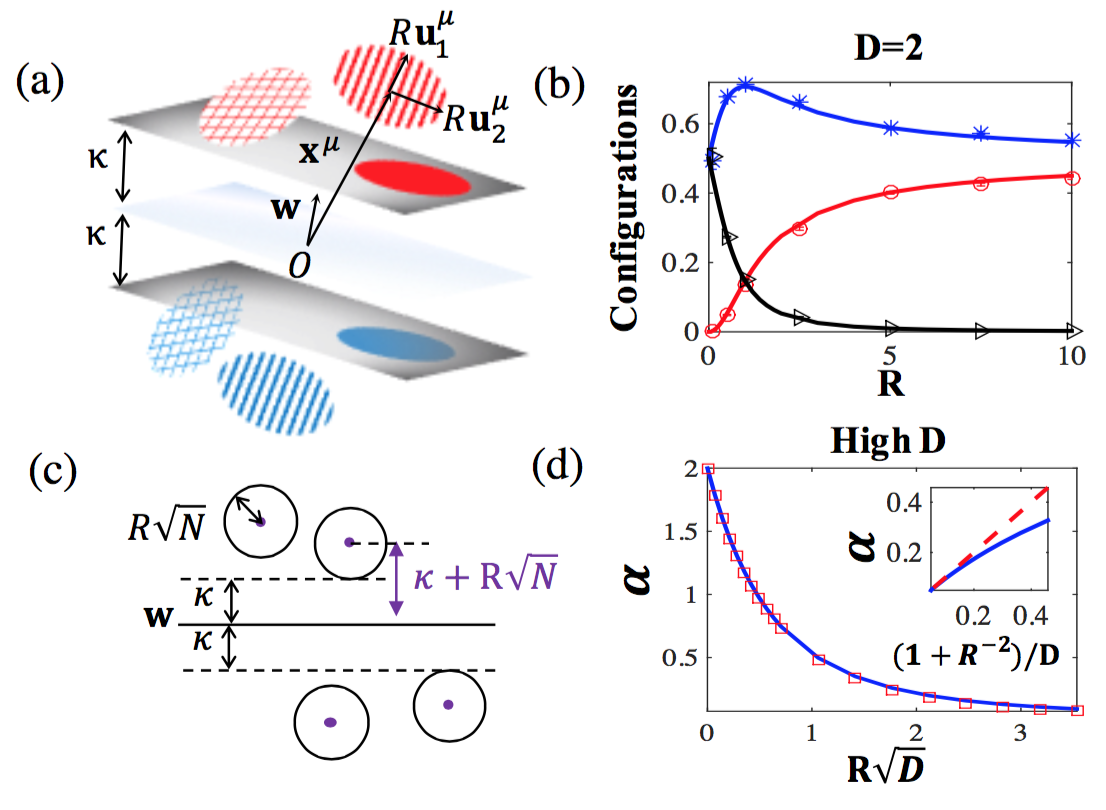} 
\par\end{centering}

\caption{Random $D$-dimensional balls: (a) Linear classification of $D=2$
balls. (b) Fraction of 2-$D$ ball configurations as a function of
$R$ at capacity with $\kappa=0$, comparing theory (lines) with simulations
(markers). (red) balls embedded in the plane, (blue) balls touching
the plane, (black) interior balls. (c) Linear classification of balls
with $D=N$ at margin $\kappa$ (black circles) is equivalent to point
classification of centers with effective margin $\kappa+R\sqrt{N}$
(purple points). (d) Capacity $\alpha=P/N$ for $\kappa=0$ for large
$D=50$ and $R\propto D^{-1/2}$ as a function of $R\sqrt{D}$. (blue
solid) $\alpha_{D}(0,R)$ compared with $\alpha_{0}(R\sqrt{D})$ (red
square). (Inset) Capacity $\alpha$ at $\kappa=0$ for $0.35\leq R\leq20$
and $D=20$: (blue) theoretical $\alpha$ compared with approximate
form $(1+R^{-2})/D$ (red dashed).\label{fig:Disks}}
\end{figure}

Higher dimensional manifolds arise from multiple sources of variability
and their nonlinear effects on the neural responses. An example is
varying stimulus orientation, resulting in two-dimensional object
manifolds under the cosine tuning function (Fig. \ref{fig:Disks}(a)).
Linear classification of these manifolds depends only upon the properties
of their convex hulls \cite{de2000computational}. We consider simple
convex hull geometries as $D$-dimensional balls embedded in $N$-dimensions:
$\left\{ \mathbf{x}^{\mu}+R\sum_{i=1}^{D}s_{i}\mathbf{u}_{i}^{\mu}\right\} $,
so that the $\mu$-th manifold is centered at the vector $\mathbf{x}^{\mu}\in\mathcal{R}^{N}$
and its extent is described by a set of $D$ basis vectors $\left\{ \mathbf{u}_{i}^{\mu}\in\mathcal{R}^{N},\:i=1,...,D\right\} $.
The points in each manifold are parameterized by the $D$-dimensional
vector $\vec{s}\in\mathcal{R}^{D}$ whose Euclidean norm is constrained
by: $\left\Vert \vec{s}\right\Vert \leq1$ and the radius of the balls
are quantified by $R$ .

Statistically, all components of $\mathbf{x}^{\mu}$ and $\mathbf{u}_{i}^{\mu}$
are i.i.d. Gaussian random variables with zero mean and unit variance.
We define $h_{0}^{\mu}=N^{-1/2}y^{\mu}\mathbf{w}\cdot\mathbf{x}^{\mu}$
as the field induced by the manifold centers and $h_{i}^{\mu}=N^{-1/2}y^{\mu}\mathbf{w}\cdot\mathbf{u}_{i}^{\mu}$
as the $D$ fields induced by each of the basis vectors and with normalization
$\left\Vert \mathbf{w}\right\Vert =\sqrt{N}$. To classify all the
points on the manifolds correctly with margin $\kappa$, $\mathbf{w}\in\mathcal{R}^{N}$
must satisfy the inequality $h_{0}^{\mu}-R||\vec{h}^{\mu}||\geq\kappa$
where $||\vec{h}^{\mu}||$ is the Euclidean norm of the $D$-dimensional
vector $\vec{h}^{\mu}$ whose components are $h_{i}^{\mu}$ . This
corresponds to the requirement that the field induced by the points
on the $\mu$-th manifold with the smallest projection on $\mathbf{w}$
be larger than the margin $\kappa$.

We solve the replica theory in the limit of $N,\,P\rightarrow\infty$
with finite $\alpha=P/N$, $D$, and $R$. The fields for each of
the manifolds can be written as sums of Gaussian quenched and entropic
components, $\left(t_{0}\in\mathcal{R},\:\vec{t}\in\mathcal{R}^{D}\right)$
and $\left(z_{0}\in\mathcal{R},\:\vec{z}\in\mathcal{R}^{D}\right)$
, respectively. The capacity for $D$-dimensional manifolds is given
by the replica symmetric calculation (SM):

\begin{equation}
\alpha_{D}^{-1}(\kappa,R)=\left\langle \min_{t_{0}+z_{0}-R\left\Vert \vec{t}+\vec{z}\right\Vert >\kappa}\frac{1}{2}\left[z_{0}^{2}+\left\Vert \vec{z}\right\Vert ^{2}\right]\right\rangle _{t_{0},\vec{t}}.
\end{equation}

The capacity calculation can be partitioned into three regimes. For
large $t_{0}>\kappa+Rt$, where $t=\left\Vert \vec{t}\right\Vert $,
$z_{0}=0$ and $\vec{z}=0$ corresponding to manifolds which lie interior
to the margin planes of the perceptron. On the other hand, when $t_{0}<\kappa-R^{-1}t$,
the minimum is obtained at $z_{0}=\kappa-t_{0}$ and $\vec{z}=-\vec{t}$
corresponding to manifolds which are fully embedded in the margin
planes. Finally, in the intermediate regime, when $\kappa-R^{-1}t<t_{0}<\kappa+Rt$,
$z_{0}=R\left\Vert \vec{t}+\vec{z}\right\Vert -t_{0}+\kappa$ but
$\vec{z}\ne-\vec{t}$ indicating that these manifolds only touch the
margin plane. Decomposing the capacity over these regimes and integrating
out the angular components, the capacity of the perceptron can be
written as: 
\begin{eqnarray}
\alpha_{D}^{-1}(\kappa,R) & = & \int_{0}^{\infty}dt\,\chi_{D}(t)\int_{\kappa-\frac{1}{R}t}^{\kappa+Rt}Dt_{0}\frac{\left(Rt+\kappa-t_{0}\right)^{2}}{R^{2}+1}\nonumber \\
 & + & \int_{0}^{\infty}dt\,\chi_{D}(t)\int_{-\infty}^{\kappa-\frac{1}{R}t}Dt_{0}\left[\left(\kappa-t_{0}\right)^{2}+t^{2}\right]\qquad\label{eq:alphaCDisks}
\end{eqnarray}

where $\chi_{D}(t)=\frac{2^{1-\frac{D}{2}}}{\Gamma(\frac{D}{2})}t^{D-1}e^{-\frac{1}{2}t^{2}}$
is the \emph{D-}Dimensional Chi probability density function. For
large $R\rightarrow\infty$, Eq. \eqref{eq:alphaCDisks} reduces to:
$\alpha_{D}^{-1}(\kappa,\infty)=\alpha_{0}^{-1}(\kappa)+D$ which
indicates that $\mathbf{w}$ must be in the null space of the $PD$
basis vectors $\left\{ \mathbf{u}_{i}^{\mu}\right\} $ in this limit.
This case is equivalent to the classification of $P$ points (the
projections of the manifold centers) by a perceptron in the $N-PD$
dimensional null space.

To probe the fields, we consider the joint distribution of the field
induced by the center, $h_{0}$, and the norm of the fields induced
by the manifold directions, $h\equiv\left\Vert \vec{h}\right\Vert $
. There are three contributions. The first term corresponds to $h_{0}-Rh>\kappa$,
i.e. balls that lie interior to the perceptron margin planes; the
second component corresponds to $h_{0}-Rh=\kappa$ but $h>0$, i.e.
balls that touch the margin planes; and the third contribution represents
the fraction of balls obeying $h_{0}=\kappa$ and $h=0$, i.e. balls
fully embedded in the margin. The dependence of these contributions
on $R$ for $D=2$ is shown in Fig. \ref{fig:Disks}(b). Interestingly,
when $\kappa=0$ , the case of $R=1$ is particularly simple for all
$D$ . The capacity is $\alpha_{D}=2/(D+1)$ ; in addition, the fraction
of embedded and interior balls are equal and the fraction of touching
balls have a maximum, see Fig. \ref{fig:Disks}(b) and SM.

In a number of realistic problems, the dimensionality $D$ of the
object manifolds could be quite large. Hence, we analyze the limit
$D\gg1$. In this situation, for the capacity to remain finite, $R$
has to be small, scaling as $R\propto D^{-\frac{1}{2}}$, and the
capacity is $\alpha_{D}(\kappa,R)\approx\alpha_{0}(\kappa+R\sqrt{D})$.
In other words, the problem of separating $P$ high dimensional balls
with margin $\kappa$ is equivalent to separating $P$ points but
with a margin $\kappa+R\sqrt{D}$. This is because when the distance
of the closest point on the $D$-dimensional ball to the margin plane
is $\kappa$, the distance of the center is $\kappa+R\sqrt{D}$ (see
Fig. \ref{fig:Disks}). When $R$ is larger, the capacity vanishes
as $\alpha_{D}(0,R)\approx\left(1+R^{-2}\right)/D$. When $D$ is
large, making $\mathbf{w}$ orthogonal to a significant fraction of
high dimensional manifolds incurs a prohibitive loss in the effective
dimensionality. Hence, in this limit, the fraction of manifolds that
lie in the margin plane is zero. Interestingly, when $R$ is sufficiently
large, $R\propto\sqrt{D}$, it becomes advantageous for $\mathbf{w}$
to be orthogonal to a finite fraction of the manifolds.

\paragraph*{$L_{p}$ balls:}

To study the effect of changing the geometrical shape of the manifolds,
we replace the Euclidean norm constraint on the manifold boundary
by a constraint on their $L_{p}$ norm. Specfically, we consider $D$-dimensional
manifolds $\left\{ x^{\mu}+R\sum_{i=1}^{D}s_{i}u_{i}^{\mu}\right\} $
where the $D$ dimensional vector $\vec{s}$ parameterizing points
on the manifolds is constrained: $\left\Vert \vec{s}\right\Vert _{p}\leq1$.
For $1<p<\infty$, these $L_{p}$ manifolds are smooth and convex.
Their linear classification by a vector $\mathbf{w}$ is determined
by the field constraints $h_{0}^{\mu}-R||\vec{h}{}^{\mu}||_{q}\geq\kappa$
where, as before, $h_{0}^{\mu}$ are the fields induced by the centers,
and $||\vec{h}^{\mu}||_{q}$, $q=p/(p-1)$, are the $L_{q}$ dual
norms of the $D$-dimensional fields induced by $\mathbf{u}_{i}^{\mu}$
(SM). The resultant solutions are qualitatively similar to what we
observed with $L_{2}$ ball manifolds.

\begin{figure}
\noindent \begin{centering}
\includegraphics[width=8.5cm]{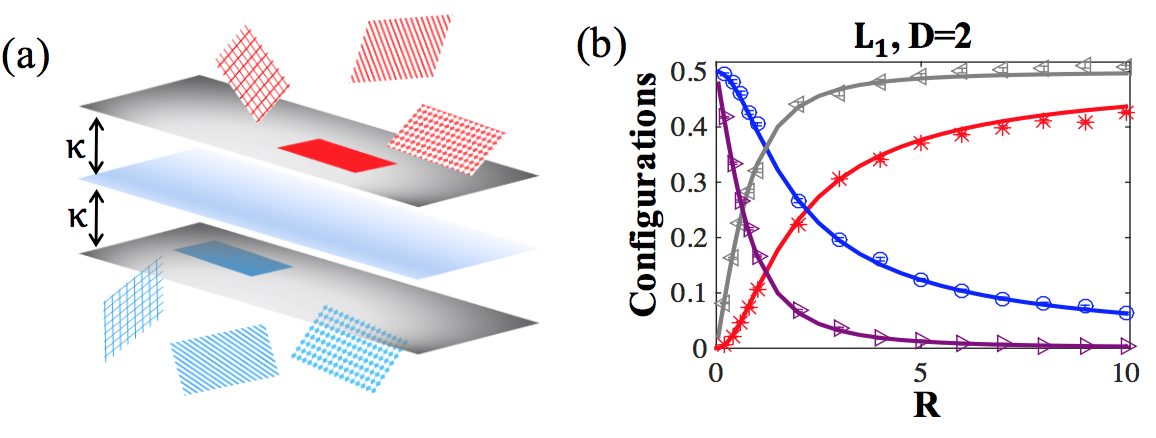} 
\par\end{centering}

\caption{$L_{1}$ balls: (a) Linear classification of 2-$D$ $L_{1}$ balls.
(b) Fraction of manifold configurations as a function of radius $R$
at capacity with $\kappa=0$ comparing theory (lines) to simulations
(markers). (red) entire manifold embedded, (blue) manifold touching
margin at a single vertex, (gray) manifold touching with two corners
(one side), (purple) interior manifold.\label{fig:L1manifolds}}
\end{figure}

However, when $p\le1$, the convex hull of the manifold becomes faceted,
consisting of vertices, flat edges and faces. For these geometries,
the constraints on the fields associated with a solution vector $\mathbf{w}$
becomes: $h_{0}^{\mu}-R\max_{i}\left|h_{i}^{\mu}\right|\ge\kappa$
for all $p<1$ . We have solved in detail the case of $D=2$ (SM).
There are four manifold classes: interior; touching the margin plane
at a single vertex point; a flat side embedded in the margin; and
fully embedded. The fractions of these classes are shown in Fig. \ref{fig:L1manifolds}.

\paragraph*{Discussion:}

We have extended Gardner's theory of the linear classification of
isolated points to the classification of continuous manifolds. Our
analysis shows how linear separability of the manifolds depends intimately
upon the dimensionality, size and shape of the convex hulls of the
manifolds. Some or all of these properties are expected to differ
at different stages in the sensory hierarchy. Thus, our theory enables
systematic analysis of the degree to which this reformatting enhances
the capacity for object classification at the higher stages of the
hierarchy.

We focused here on the classification of fully observed manifolds
and have not addressed the problem of generalization from finite input
sampling of the manifolds. Nevertheless, our results about the properties
of maximum margin solutions can be readily utilized to estimate generalization
from finite samples. The current theory can be extended in several
important ways. Additional geometric features can be incorporated,
such as non-uniform radii for the manifolds as well as heteogeneous
mixtures of manifolds. The influence of correlations in the structure
of the manifolds as well as the effect of sparse labels can also be
considered. The present work lays the groundwork for a computational
theory of neuronal processing of objects, providing quantitative measures
for assessing the properties of representations in biological and
artificial neural networks. 
\begin{acknowledgments}
Helpful discussions with Remi Monasson and Uri Cohen are acknowledged.
The work is partially supported by the Gatsby Charitable Foundation,
the Swartz Foundation, the Simons Foundation (SCGB Grant No. 325207),
the NIH, and the Human Frontier Science Program (Project RGP0015/2013).
D. Lee also acknowledges the support of the US National Science Foundation,
Army Research Laboratory, Office of Naval Research, Air Force Office
of Scientific Research, and Department of Transportation.\end{acknowledgments}

\end{document}